\documentclass[aps,pra,twocolumn,superscriptaddress,showpacs]{revtex4}
\usepackage{amssymb}

\usepackage{graphicx}
\usepackage{dcolumn}
\usepackage{bm}

\begin{document}

\title{Assessment of density functional approximations: long-range correlations and
self-interaction effects}

\author{J. Jung}
\affiliation{ Departamento de F\'{\i}sica Fundamental, Universidad Nacional
              de Educaci\'{o}n a Distancia, Apartado 60141, E-28080
              Madrid, Spain }
 
\author{P. Garc\'{\i}a-Gonz\'{a}lez}
 
\affiliation{ Departamento de F\'{\i}sica de la Materia Condensada, C-III,
              Universidad Aut\'{o}noma de Madrid,
              E-28049 Madrid, Spain }
 
\author{J.~E. Alvarellos}
\affiliation{ Departamento de F\'{\i}sica Fundamental, Universidad Nacional
              de Educaci\'{o}n a Distancia, Apartado 60141, E-28080
              Madrid, Spain }

\author{R.~W. Godby}
\affiliation{ Department of Physics, University of York, Heslington,
              York YO10 5DD, United Kingdom }

\begin{abstract}
The complex nature of electron-electron correlations is made manifest in the
very simple but non-trivial problem of two electrons confined within a
sphere. The description of highly non-local correlation and self-interaction
effects by widely used local and semi-local exchange-correlation energy
density functionals is shown to be unsatisfactory in most cases. 
Even the best such
functionals exhibit significant errors in the Kohn-Sham potentials and density
profiles.
\end{abstract}

\pacs{31.15.Ew,31.25.-v,71.15.Mb}
\maketitle

\section{Introduction}

The Kohn-Sham (KS) \cite{KS} formulation of Density Functional Theory (DFT) 
\cite{HK} is in the present day the most popular method in electronic
structure calculations. In this scheme, the exact ground state energy and
electron density could be found self-consistently if the
exchange-correlation (XC) energy functional $E_{\mathrm{XC}}[n]$ was known. $%
E_{\mathrm{XC}}[n]$ contains all the quantum many-body effects of the
electron system, but very simple mean-field prescriptions like the Local
Density Approximation (LDA) often suffice to obtain accurate results for a
wide variety of systems at an affordable computational cost.

However, there are several problems that are well beyond the capabilities of
the local approximation and any \textit{semi-local} extension thereof,
because of the clear manifestation of the very non-local nature of
electron-electron correlations. For example, the long-ranged van der Waals
interactions, the image potential at metal surfaces and clusters, or several 
\textit{pathological} behaviors of the exact XC potential cannot be
described at all using simple XC functional models \cite{GG}. Nevertheless,
these limitations should not be important in many other situations, and new
XC functionals are being proposed with the aim of reaching chemical accuracy
while keeping the implementation ease of the local density based
approximations \cite{Na,PK}. This could open the appealing possibility of
making predictive studies of relevant aspects of Quantum Chemistry such as
reaction paths, atomization energies, and bond lengths and energies.

The purpose of this paper is to bring further insight in the capabilities
and limitations of local and semi-local XC functionals, showing that even in
very simple problems the complexity of the quantum electron-electron
correlations might prevent any of these approaches from properly describing
the ground state properties of such systems. We do not intend to make a
comprehensive assessment of mean-field-like approximations, but just to
provide a representative common picture of all them. Hence, among the realm
of proposals existing in the literature, we have chosen the well-known LDA
prescription by Perdew and Wang \cite{LDA}, the Generalized Gradient
Approximation (GGA) by Perdew-Burke-Erzenhof \cite{PBE}, and the very recent meta-GGA (MGGA) proposed
by Tao \textit{et al} \cite{Tao}. These three approaches have the virtue of
being designed using general considerations (i.e. they do not include
empirical parameters). Furthermore, 
they can be seen as a coherent set of conceptual
progressive improvements starting from the strictly local
approximation, then considering the dependence on the density variation, and
finally including information from KS orbitals through its associated kinetic
energy density.

\begin{table*}[t!]
\caption{Comparison between the exact exchange and correlation energies for
several sphere radii $R$ and the results given by the local and semi-local
functionals evaluated on the exact density profile. The exact total energy
is also included to illustrate the increasing importance of correlation for
high $R$. }
\label{Tab1}%
\begin{ruledtabular}
\begin{tabular}{rr|rrrr|rrrr}
\multicolumn{1}{c} {$R$}                       & 
\multicolumn{1}{c|}{$E_{\rm tot}^{\rm ex}$}    &
\multicolumn{1}{c} {$E_{\rm X}^{\rm ex}$}      &
\multicolumn{1}{c} {$E_{\rm X}^{\rm LDA}$}     & 
\multicolumn{1}{c} {$E_{\rm X}^{\rm GGA}$}     & 
\multicolumn{1}{c|}{$E_{\rm X}^{\rm MGGA}$}    & 
\multicolumn{1}{c} {$E_{\rm C}^{\rm ex}$}   & 
\multicolumn{1}{c} {$E_{\rm C}^{\rm LDA}$}     & 
\multicolumn{1}{c} {$E_{\rm C}^{\rm GGA}$}     & 
\multicolumn{1}{c} {$E_{\rm C}^{\rm MGGA}$}   \\ \hline
1 & 11.5910 & -1.7581 & -1.5230 & -1.6843 & -1.7805
            & -0.0507 & -0.1424 & -0.0775 & -0.0647 \\ 
5 & 0.7016  & -0.3335 & -0.2914 & -0.3213 & -0.3412 
            & -0.0383 & -0.0715 & -0.0501 & -0.0421 \\
10 & 0.2381 & -0.1592 & -0.1407 & -0.1550 & -0.1651
            & -0.0288 & -0.0481 & -0.0362 & -0.0301 \\
25 & 0.0633 & -0.0590 & -0.0537 & -0.0596 & -0.0634 
            & -0.0163 & -0.0257 & -0.0201 & -0.0166 \\
50 & 0.0249 & -0.0278 & -0.0262 & -0.0296 & -0.0311
            & -0.0093 & -0.0151 & -0.0116 & -0.0095
\end{tabular}
\end{ruledtabular}
\end{table*}

We will study a very simple but non-trivial system: two electrons confined
within a sphere of hard walls, whose solution has been found through
accurate Configuration Interaction calculations \cite{JA,TA}. This system is
an interesting benchmark reference due to the following reasons. First, its
simplicity: the density is isotropic, hence having a simple mathematical
one-dimensional problem restricted to the radial coordinate. Second, its
ground state has singlet spin configuration. As a consequence, Pauli's
correlation (exchange) between the electrons is absent and the exchange
energy $E_{\mathrm{X}} \left[ n\right] $ just corrects the spurious electron
self-interaction in the classical Hartree electrostatic energy $W_{\mathrm{H}%
}\left[ n\right]$. That is, the Coulomb correlation is actually the only
source of quantum many-body effects in this system. Finally, different
correlation regimes can be easily achieved by varying the radius $R$ of the
confining sphere. Thus, at small $R$ (high mean density) we are in the
low-correlation limit where the confinement by the sphere dominates over the
electron-electron interaction, so having a system with an \textit{atomic-like%
} behavior. By increasing the value of $R$ (decreasing the mean density) we
gradually enter into a highly correlated regime in which the correlation
exhibits long-ranged and anisotropic effects.

Then, this simple system offers an excellent scenario to assess essential
features of functional approximations to the XC energy. In particular, since
it is impossible within the present formulation of semi-local functionals to
achieve the exact exchange energy for arbitrary one- and two- electron
systems ($-W_{\mathrm{H}}\left[ n\right] $ and $-W_{\mathrm{H}}\left[ n%
\right] /2$ respectively), we can easily see how important is this
limitation for two-electron densities with very distinct mean densities. On
the other hand, LDA and GGA suffer from spurious correlation self-interaction, 
a limitation which is corrected by the MGGA \cite{Tao,PKZB}. 
Nonetheless, the proper self-interaction
correction does not guarantee the overall accuracy of the correlation
functional, and the actual role played by non-local correlation effects
has to be checked carefully.

As a first test, we will evaluate these functionals over the exact
densities, i.e. in a non-self-consistent fashion, comparing the XC energies
and potentials with the exact ones. Then, we will present the fully
self-consistent solutions, in such a way that we will assess not only the
self-consistentent energies but also the DFT densities that minimizes the
corresponding total energy functionals. Atomic units ($\hbar =m_{e}=e=1$)
will be used throughout the paper.

\section{Results on the exact density profile}

The Hamiltonian of the two-electron model system is given by 
\begin{eqnarray*}
\hat{H}=-\frac{1}{2}\sum_{i=1}^{2}\nabla _{i}^{2} &+&\frac{1}{\mid \vec{r}%
_{1}-\vec{r}_{2}\mid }+\sum_{i=1}^{2}V\left( r_{i}\right) \\
&& \\
V\left( r\right) &=&\left\{ 
\begin{array}{l@{\quad \quad}l}
0 & \quad r<R \\ 
\infty & \quad r\geq R.
\end{array}
\right.
\end{eqnarray*}
The hard wall described by $V(r)$ impose strict boundary conditions at $r=R$%
, but does not have a direct contribution to the total energy functional $E%
\left[ n\right] $. Therefore: 
\begin{equation}
E[n]=T_{\mathrm{S}}[n]+W_{\mathrm{H}}[n]+E_{\mathrm{XC}}[n]~,  \label{energy}
\end{equation}
where $T_{\mathrm{S}}[n]$ is the kinetic energy of the fictitious KS
non-interacting system. For a singlet state, the set of KS equations is
reduced to a single Schr\"{o}dinger equation 
\begin{equation}
\left[ -\frac{1}{2}\nabla ^{2}+v_{\mathrm{S}}(\mathbf{r)}\right] \phi (%
\mathbf{r})=\varepsilon \phi (\mathbf{r})~,  \label{Schrod}
\end{equation}
where $v_{\mathrm{S}}=v_{\mathrm{H}}+v_{\mathrm{X}}+v_{\mathrm{C}}$ is the
KS effective potential which is the sum of the Hartree $(v_{\mathrm{H}})$,
exchange $(v_{\mathrm{X}})$, and correlation $(v_{\mathrm{C}})$ potentials.
The density is related to the ground-state orbital of (\ref{Schrod}) through
the simple equality $\phi (\mathbf{r})=\sqrt{n(\mathbf{r})/2}$.

Under the exact DFT formulation, if $n^{\mathrm{ex}}(r)$ is the ground-state
density of the system, the corresponding eigenvalue $\varepsilon ^{\mathrm{ex%
}}$ must equal the ionization energy $E[n^{\mathrm{ex}}]-E^{(1)}$, where $%
E[n^{\mathrm{ex}}]=E_{\mathrm{tot}}^{\mathrm{ex}}$ is the two-electron
ground state energy and $E^{(1)}=\pi ^{2}/\left( 2R^{2}\right) $ is the
energy of the one-electron system. Thus, the exact correlation potential can
be written explicitly as 
\begin{equation}
v_{\mathrm{C}}^{\mathrm{ex}}(\mathbf{r})=\varepsilon ^{\mathrm{ex}}+\frac{1}{%
2}\frac{\nabla ^{2}\sqrt{n^{\mathrm{ex}}(\mathbf{r})}}{\sqrt{n^{\mathrm{ex}}(%
\mathbf{r})}}-\frac{1}{2}\int d\mathbf{r}^{\prime }\frac{n^{\mathrm{ex}}(%
\mathbf{r})}{\left| \mathbf{r}-\mathbf{r}^{\prime }\right| },  \label{potef}
\end{equation}
where we have used the exact relation $E_{\mathrm{X}}\left[ n\right] =-W_{%
\mathrm{H}}\left[ n\right] /2$.\cite{note1} However, 
it is worth pointing out that (\ref{potef}) is an expression only 
valid for the exact density profile. 
An exact functional expression for the correlation
potential $v_{\mathrm{C}}\left( \mathbf{r}\right) $
of an arbitrary spin-unpolarized two-electron density $n\left( 
\mathbf{r}\right) $ would require to know the external
potential that defines the two interacting electron system whose ground
state is $n(\mathbf{r})$ and then include such a potential.
Then, the simplicity suggested by (\ref{potef}) is just apparent.

\begin{figure}[t!]
\includegraphics[width=8.4cm,bb=78 135 475 681,clip]{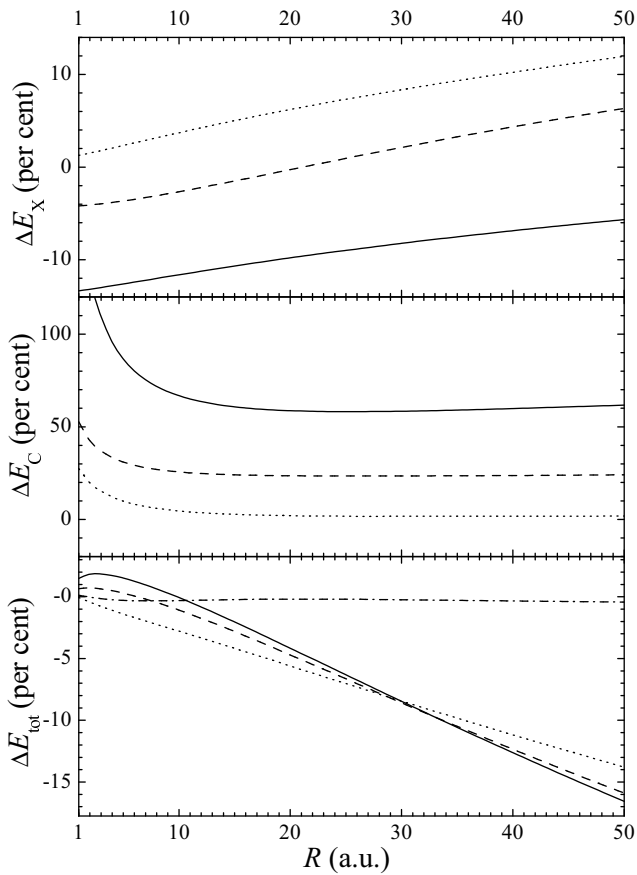}
\caption{Percent errors $\Delta E = 100\times\left(E^{\mathrm{DFT}}-E^{%
\mathrm{ex}}\right)/E^{\mathrm{ex}}$ for the exchange, correlation, and
total energies as functions of the sphere radius $R$. All DFT results are
obtained non-self-consistently over the exact density. Solid line: LDA;
dashed line: GGA; dotted line: MGGA; dash-dotted line: EXX+M.}
\label{Fig1}
\end{figure}

The performance of different local and semi-local prescriptions when
evaluating the XC energy on the exact density profile is shown in Fig. \ref
{Fig1} and Table \ref{Tab1}. The LDA systematically underestimates the
absolute value of the exchange energy, whereas the GGA partially corrects
this trend, although in the low density limit the GGA overestimates $|E_{%
\mathrm{X}}|$. The MGGA behaves reasonably well for small radii, which is
not a surprise since it reproduces exactly the exchange energy of the
hydrogen atom and, as we said in the Introduction, in this range the model
system behaves precisely like an atom. Thus, although the MGGA does not
cancel exactly the spurious Hartree self-interactions, it fairly accounts
for such a cancellation in atomic-like systems. When decreasing the mean
electron density, the system cannot be considered like an atom any more and the
MGGA greatly overestimates the self-interaction corrections to $W_{\mathrm{H}%
}$, becoming even worse than GGA.

Regarding the correlation energy $E_{\mathrm{C}}\left[ n\right] $, the LDA
shows an evident poor behavior which is improved by the GGA although the
correlation energies are always much too negative. On the contrary, the MGGA
behaves extremely well for all densities. Its relative error in the high
density limit (around 20\%) has a minor influence in the total energy, and
such an error is less than 5\% for lower densities. This excellent
performance is in agreement with the conclusions by Seidl \textit{et al} 
\cite{SPK} about the essentially correct behavior of the meta-GGA
correlation functional proposed in Ref. \onlinecite{PKZB} 
(the basic ingredient of the MGGA by Tao 
\textit{et al}) under uniform scaling to the low density limit. We have to
bear in mind that in the highly correlated regime, the two electrons are 
\textit{localized} in two different positions. That is, if one of the
electrons is at a distance $r$ from the center, the probability to find the
second one is concentrated around a point on the opposite side \cite{JA}.
Thus, the LDA/GGA main source of error in this regime, corrected by the
MGGA, is the absence of self-interaction corrections.

In the lower panel of Fig. \ref{Fig1} we present the accuracy of the
corresponding non-self-consistent DFT total energies. The well known
compensation of errors between exchange and correlation in LDA and GGA is
easy to observe in the atomic-like limit, but in the high correlation regime
both functionals fail badly. The MGGA does not benefit from any cancellation
of errors (it overestimates the absolute value of exchange \textit{and}
correlation energies). Hence, it slighty underestimates the total energy for
small radii, but the error on exchange dominates for lower densities where
the MGGA perfoms poorly. A hyper-GGA \cite{PK}, designed with the aim of
being free of any self-interaction error, must yield very accurate total
energies for all the ranges in our model system. Then, its performance
should be similar to the presented in the lower panel of Fig. \ref{Fig1},
where we plot the relative error on the total energy evaluated through a
hybrid functional, EXX+M, where the exchange is calculated exactly and the
correlation approximated by the MGGA. 
In this case, the error on the total
energy is always less than 1 \%, and for $R\simeq 1$ the absolute deviation
is just 7 mHa/e, fairly close to the chemical accuracy (around 2 mHa/e).

\begin{figure}[t!]
\includegraphics[width=8.4cm,bb=65 140 475 683,clip]{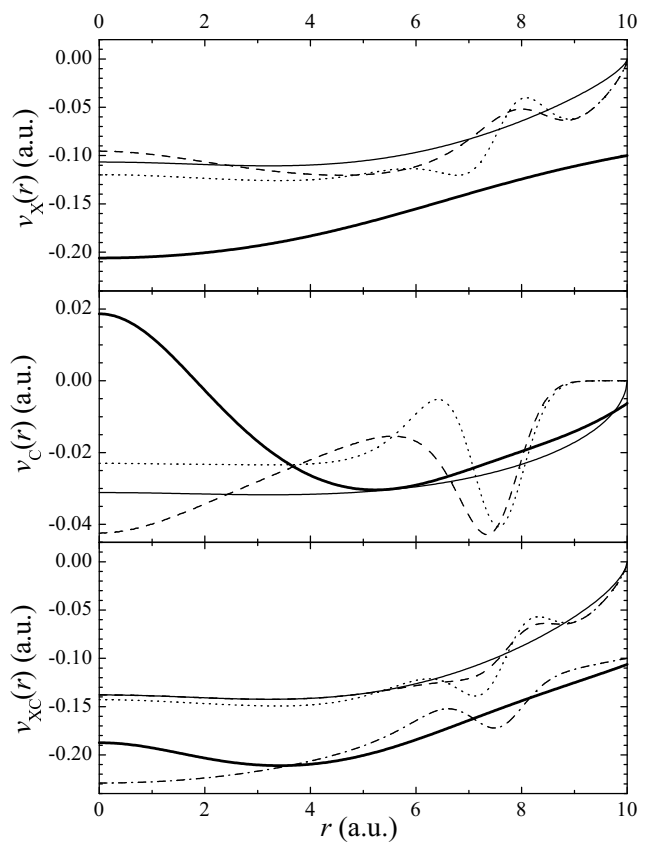}
\caption{Exchange and correlation potentials obtained from the exact density
profile ($R = 10$). Thick solid line: exact results; thin solid line: LDA;
dashed line: GGA; dotted line: MGGA; dash-dotted line: EXX+M. 
Note that the
shape of the LDA XC potential reproduces fairly well the exact one excepting
in the region around the center of the sphere. This error is amplified by
the EXX+M prescription, where the exchange part is exactly correct but 
the correlation potential is the same as the MGGA.}
\label{Fig2}
\end{figure}

From the shape of the potentials it is possible 
to see the underlying physics contained in the approximations.
The above mentioned \textit{localization} of the electrons in the highly
correlated limit is an obvious consequence of the electrostatic repulsion,
which tend to dominate over the confinement by the wall as we increase the
radius of the system. Since the Hartree energy contains spurious
self-interactions, the role of the exchange is to compensate
partially the effects due to $W_{\mathrm{H}}$. As we can see in the upper
panel of Fig. \ref{Fig2}, where $v_{\mathrm{X}}(r)$ is plotted for the model
system with $R=10$, the exact exchange potential has a mininum in the center
of the sphere, favouring an atomic-like behavior. The corresponding LDA
potential is not able to account completely for this exchange \textit{%
attractive} feature, whereas the overall shift of $v_{\mathrm{X}}^{\mathrm{%
LDA}}$ is a concomitant consequence of the local dependence on the density.
The semi-local potentials exhibit a similar behavior but there are
unphysical oscillations reflecting the presence of the gradient and the
Laplacian of the density in the expression of the exchange potential.

On the contrary, the Coulomb correlation enhances the localization through a
potential barrier located in the center of the sphere (see the middle panel
of Fig. \ref{Fig2}). This barrier reflects a truly non-local correlation
effect. In fact, for $R=10$, the electron density is almost 
homogeneous around $r=0$, 
and then $v_{\rm C}^{\rm LDA}({\bf r})$ is practically constant in this 
region. 
On the other hand, the LDA fits reasonably well the exact potential 
if $r\gtrsim 5$, 
but this partial agreement is completely lost under the GGA and MGGA. 
This overall
bad quality of the local and semi-local correlation potentials is a
general feature in finite systems \cite{UG}, although in some cases it could be
masked if we focus on the total $v_{\rm XC}({\bf r})$. 
As we may see in the lower panel of Fig. \ref{Fig2}, the XC
potentials given by LDA, GGA, and MGGA are rather similar (excepting
the above mentioned unphysical oscillations). Their shape is close to
the exact one for $r\gtrsim 5$ but, as expected, they do not
reproduce at all the correlation barrier at $r=0$.  
Under the EXX+M prescription, 
the situation is even worse, since the XC potential reaches a minimum 
at $r=0$ due to an exact description of exchange which is not compensated 
by an accurate
correlation potential. Therefore, although the correlation energies 
given by the MGGA are excellent, 
the potentials derived from it
are not able to reproduce the non-local effects that
manifest themselves in the shape of $v_{\mathrm{C}}(\mathbf{r})$. As we will
see in the next section, this will lead to important deviations from
the exact density profile when solving self-consistently the KS equation.

\section{Self consistent results}

One of the major advantages of KS-DFT is its fully self-consistent
character: any previous knowledge of the electron density profile is not
required excepting for setting up an initial guess to start the iterative
resolution of the KS equations. For many purposes, LDA and/or GGA give
accurate self-consistent densities because the corresponding approximate
potentials are very similar to the exact ones in those regions relevant for
the calculation of the electron density. This justifies the use of more
sophisticated functional expressions as a mere correction over the
self-consistent LDA/GGA densities. Consequently, the main effort carried
out during the last years had been directed towards the improvement of the
XC energies, paying less attention to the characteristics of the XC
potential $v_{\mathrm{XC}}(\mathbf{r})$.

\begin{figure}[t!]
\includegraphics[width=8.4cm,bb=65 279 475 531,clip]{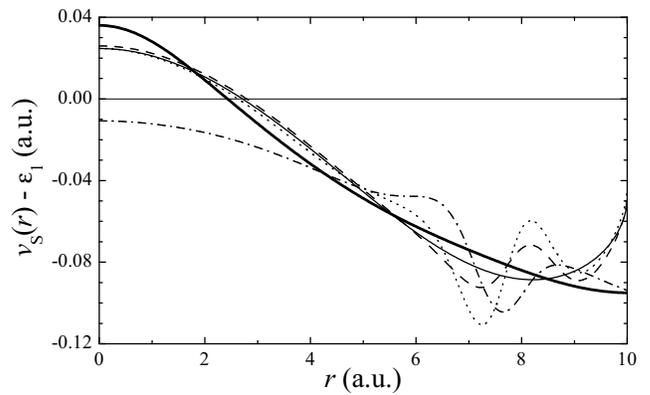}
\caption{Kohn-Sham potential $v_{\mathrm{S}} \left( r \right)$ for the exact
density profile ($R = 10$). Thick solid line: exact result; thin solid line:
LDA; dashed line: GGA; dotted line: MGGA; dash-dotted line: EXX+M. 
In order to allow an easier comparison, the
potentials have been shifted in such a way that the corresponding first
eigenvalue for each approximate potential is equal to zero.
The behavior
of the potentials near the sphere walls is less important due to the
role played by the boundary condition $n(R)=0$.}
\label{Fig3}
\end{figure}

Nonetheless, in the previous section we have seen that all the functionals
considered in this paper fail to reproduce the exact XC potential in a
region where the electron density is far from being negligible. The
consequences can be seen in Fig. \ref{Fig3}, where we compare the exact KS
potential $v_{\mathrm{S}}(r)$ with the DFT ones obtained from the exact
density profile $n^{\mathrm{ex}}\left( r\right) $. The LDA, GGA, and MGGA
do not reproduce the exact shape of $v_{\mathrm{S}}(r)$ and it is
reflected by a classical forbidden region greater than the actual one: the
self-consistent density will be pushed towards the walls. The EXX+M model,
as comented, incorporates exactly the attractive character of exchange, but
the wrong description of $v_{\mathrm{C}}(r)$ makes the effective potential
less confining than the exact one: this hybrid approach tends to concentrate
the density around $r=0$.

\begin{figure}[t!]
\includegraphics[width=8.4cm,bb=90 227 514 596,clip]{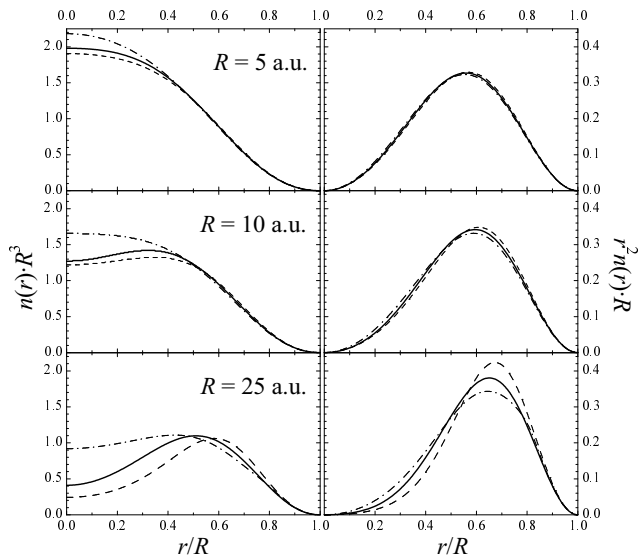}
\caption{Exact (solid line) and self-consistent GGA (dashes) and EXX+M
(dash-dots) densities $n\left( r \right)$ and reduced radial densities $r^2
n\left( r \right)$. All the quantities have been scaled with the radius of
the confining sphere. The self-consistent LDA and MGGA densities are very
similar to the GGA ones and have not been included in the figure. None of 
the approximate functionals is able to reproduce the
correct behavior of the density in the center of the sphere, and their
overall performance is very poor for $R \gg 0$.}
\label{Fig4}
\end{figure}

These deviations from the exact two-electron density profile, which can be
quantified through the expression 
\begin{equation}
\delta n=\frac{1}{2}\int \mathrm{d}\mathbf{r}\left| n^{\mathrm{ex}}(\mathbf{r%
})-n(\mathbf{r})\right| \,,  \label{nerror}
\end{equation}
should not be important for high mean densities. In this atomic-like limit
the external confining potential dominates, and the electrons are going to
be concentrated around the center of the sphere anyway. However, the greater
the radius the less important the confinement, and the shape of the XC
potential will play a more prominent role. This trend can be seen in Fig. 
\ref{Fig4}, where we compare the DFT densities $n(r)$ as well as the the
corresponding reduced radial densities $r^{2}n(r)$ with their exact
counterparts. In the atomic-like regime ($R\lesssim 5$), we can observe
genuine errors on the DFT densities around the center of the sphere.
However, this wrong behavior will lead to marginal errors on integrated
quantities, as suggested by the overall good agreement shown by the reduced
radial densities. At intermediate mean densities ($R\simeq 10$), the
differences on $r^{2}n(r)$ can be already observed at a first glance.
Moreover whereas the system shows an incipent localized behavior, which is
characterized by a density reaching a local maximum at $r\neq 0$, the EXX+M
self-consistent density still has an atomic behavior characterized by a
maximum at $r=0$. Finally, in the low density limit, all the approximate
functionals fail to describe the exact density profile with a minimum
accuracy. For instance, if $R=25$ the error given by (\ref{nerror}) is 8.6
\% using the EXX+M functional and 14.2 \% using the GGA.

\begin{table}[t!]
\caption{Self-consistent DFT-KS results for several radii compared with the
exact ones. The last file of each entry contains the DFT density error $%
\protect\delta n$ as defined in Eq. \ref{nerror}.}
\label{Tab2}%
\begin{ruledtabular}
\begin{tabular}{l|rrrrr}
 & \multicolumn{1}{c}{Exact} & \multicolumn{1}{c}{LDA}  
 & \multicolumn{1}{c}{GGA}   & \multicolumn{1}{c}{MGGA}     
 & \multicolumn{1}{c}{EXX+M} \\
\hline 
$R=1$           &&&&& \\
$E_{\rm tot}$         & 11.5910 & 11.7338 & 11.6376 & 11.5540 & 11.5770 \\
$T_{\rm S}+W_{\rm H}$ & 13.3999 & 13.3955 & 13.3974 & 13.3973 & 13.4000 \\
$E_{\rm X}$           & -1.7581 & -1.5193 & -1.6820 & -1.7785 & -1.7582 \\
$E_{\rm C}$           & -0.0507 & -0.1423 & -0.0778 & -0.0648 & -0.0647 \\
$\delta n$            &         &  0.0070 &  0.0049 &  0.0056 &  0.0006 \\
\hline 
$R=5$           &&&&&  \\
$E_{\rm tot}$         &  0.7016 &  0.7104 &  0.7017 &  0.6897 &  0.6975 \\
$T_{\rm S}+W_{\rm H}$ &  1.0734 &  1.0713 &  1.0716 &  1.0720 &  1.0747 \\
$E_{\rm X}$           & -0.3335 & -0.2897 & -0.3196 & -0.3401 & -0.3348 \\
$E_{\rm C}$           & -0.0383 & -0.0713 & -0.0504 & -0.0421 & -0.0423 \\
$\delta n$            &         &  0.0210 &  0.0222 &  0.0196 &  0.0157 \\
\hline
$R=10$          &&&&&  \\
$E_{\rm tot}$         &  0.2381 &  0.2371 &  0.2346 &  0.2305 &  0.2362 \\
$T_{\rm S}+W_{\rm H}$ &  0.4261 &  0.4245 &  0.4246 &  0.4249 &  0.4275 \\
$E_{\rm X}$           & -0.1592 & -0.1395 & -0.1537 & -0.1643 & -0.1608 \\
$E_{\rm C}$           & -0.0288 & -0.0479 & -0.0363 & -0.0301 & -0.0306 \\
$\delta n$            &         &  0.0383 &  0.0448 &  0.0371 &  0.0401 \\
\hline
$R=25$          &&&&&  \\
$E_{\rm tot}$         &  0.0633 &  0.0587 &  0.0584 &  0.0582 &  0.0626 \\
$T_{\rm S}+W_{\rm H}$ &  0.1387 &  0.1377 &  0.1376 &  0.1377 &  0.1395 \\
$E_{\rm X}$           & -0.0590 & -0.0533 & -0.0597 & -0.0632 & -0.0600 \\
$E_{\rm C}$           & -0.0163 & -0.0256 & -0.0195 & -0.0163 & -0.0170 \\     
$\delta n$            &         &  0.1352 &  0.1424 &  0.1300 &  0.0858 \\
\hline
$R=50$      &&&&&  \\
$E_{\rm tot}$         &  0.0249 &  0.0197 &  0.0199 &  0.0204 &  0.0243 \\
$T_{\rm S}+W_{\rm H}$ &  0.0620 &  0.0621 &  0.0622 &  0.0623 &  0.0624 \\
$E_{\rm X}$           & -0.0278 & -0.0270 & -0.0318 & -0.0330 & -0.0282 \\
$E_{\rm C}$           & -0.0093 & -0.0154 & -0.0105 & -0.0088 & -0.0099 \\    
$\delta n$            &         &  0.3317 &  0.3348 &  0.3343 &  0.1084 \\
\end{tabular}
\end{ruledtabular}
\end{table}
Then, the full minimizimation of $E[n]$ adds a further source of error
due to the inaccuracies of the self-consistent density. Nonetheless, these
{\it self-consistency-induced} errors in the total energies 
are going to be less important because 
there is an overall trend to cancellation, see Table \ref{Tab2}. 
For instance, the EXX+M self-consistent density is smoother than the exact 
one, which lowers the kinetic and exchange energies while increasing 
the Hartree interaction energy. However, in spite of the distorted density 
profile, there is a fortunate cancellation of errors that makes the sum
of these three terms practically equal to the exact value. Thus, the
minimization of the energy leads to changes on the density profile favoring
lower correlation energies, which is done by increasing the density around the
center of the sphere. As a result, self-consistent total energies 
are just slighty worse than the non-self-consistent ones, the
relative error $\Delta E_{\mathrm{tot}}$ ranging from -0.1\% for $R=1$ to
-2.5\% for $R=50$. 
A similar conclusion reads for the remaining functionals 
(LDA, GGA, MGGA), although in this case the self-consistent $T_{\rm S}[n]$ 
is greater than the exact one, whereas the classical interaction energy 
is reduced after the minimization. Hence, self-consistency keeps the fairly 
good quality of the total energies in the atomic limit, where there are minor
changes in the density profile, and for intermediate and low densities the 
dramatic changes on $n({\bf r})$ induce a few percent variation on the 
total energies that, in any case, were already too small.

\section{Conclusions}
In this work we have assessed the quality of some of the most popular
XC functionals used in \textit{ab-initio} electronic structure calculations
in a simple system of electrons where several correlation
regimes can be easily achieved. None of the exchange functionals is
completely free of self-interaction errors, making imposible to obtain
accurate energies for highly correlated electrons. Local and GGA correlation
exhibit a similar drawback, but the MGGA correlation shows an excellent
performance for all the regimes. Nonetheless, neither LDA/GGA nor MGGA can
describe highly non-local correlation effects that lead to a non-trivial
behavior of the correlation potential, so seriously affecting the quality 
of the
self-consistent densities. Due to persistent cancellation of trends, the
corresponding changes on the total energy after minimization are less
important, but this uncontrolled source of error might prevent these 
approximate functionals from having full predictive accuracy. 

The overall bad quality of the self-consistent KS potential also compromises
the evaluation of post-self-consistency corrections based on more
sophisticated methods, like Many-Body Perturbation Theory or time-dependent
DFT \cite{GG}, if the wrong effective potential is not corrected as well. 
On the other hand, an accurate description of the XC
potential is required, for instance, when studying neutral excitations in
finite systems using time-dependent DFT \cite{ORR}. 
As shown recently by Della Sala and G\"{o}rling \cite{Gor}, 
for those systems having an HOMO orbital 
with nodal surfaces, the exact exchange potential tends to a constant if going 
to infinity over a set of zero measure directions. This leads to
the appearance of potential barriers that, although could be
of minor importance when obtaining the self-consistent static results, might
be essential if a proper description of all the unoccupied KS orbitals was
required. Here, although in a very different context, potential barriers
induced by the non-locality of the many-body effects in an electron system
have been observed as well.

The limitations observed for this
family of prototype systems might have relevance for real molecular or
condensed-matter systems in some cases.
Prospective tests of the MGGA functional used in this work show 
an excellent performance for a wide variety of typical molecular and solid 
state systems which, moreover, seems to be kept if partial 
self-consistency is achieved \cite{Tao}. However, there is no guarantee 
that the energy minimization procedure may lead to small, but relevant changes 
on the local electron density in, for instance, the bonding region between 
two species, so giving a wrong account of the nature of such a chemical bond. 
Finally, for this model system we have seen that there are not substantial 
differences between the fully self-consistent GGA and MGGA densities, 
although in this case the MGGA functionals take a simple GGA-like form. 
In spite of this simplification, the MGGA-XC potential
amplifies the spurious oscillations already appearing under the GGA
prescription. As a conclusion, the overall capability of MGGA and envisaged
improvements thereof to yield accurate XC energies can hardly be 
questioned (specially for the correlation part), but further studies
of the actual performance of the corresponding KS potentials are required.
The latter point is relevant for those situations in which LDA/GGA are not
able to reproduce the electron density with the requiered accuracy, and
for other DFT-based applications needing an overall good description of
$v_{\mathrm S}({\mathbf r})$.
\begin{acknowledgments}
We wish to thank R. Almeida, T. Gould, J.~M. Soler, D.~Garc\'{\i}a Aldea, J. Tao 
and D.~C. Thompson for
fruitful discussions, and K. Burke for providing the subroutine for 
the PBE GGA functional. This work has been funded in part by the EU through the
NANOPHASE Research Training Network (Contract No.\ HPRN-CT-2000-00167),
the Spanish Science and Technology Ministry grant BFM2001-1679-C03-03
and by a grant from the UNED Fund for Research 2001.
\end{acknowledgments}


\begin{thebibliography}{99}
\bibitem{note1} Note that in this particular case with 
two electrons the exact exchange energy defined 
in the KS-DFT formalim is equivalent to the usual Restricted 
Hartree Fock exchange energy.
\end{thebibliography}
\end{document}